\documentclass[aps,prl,twocolumn,showpacs,%
    amsmath,%
    reprint%
    ,preprintnumbers,
    ]{revtex4-1}

\usepackage{amsmath,amssymb}
\usepackage{graphicx}
\usepackage{enumerate}

\newcommand{\Z}{\mathbb{Z}}
\newcommand{\R}{\mathbb{R}}

\newcommand{\N}{\mathbb{N}}
\newcommand{\cS}{\mathcal{S}}
\newcommand{\D}{\mathrm{d}}
\newcommand{\e}{\mathrm{e}}

\begin{document}

\title{Quantum graphs with the Bethe--Sommerfeld property}

\author{Pavel Exner$^{\*}$}
\email[]{exner@ujf.cas.cz}
\homepage[]{http://gemma.ujf.cas.cz/~exner/} \affiliation{Doppler
Institute for Mathematical Physics and Applied Mathematics, Czech
Technical University, B\v rehov{\'a} 7, 11519 Prague, Czechia,}
\affiliation{Department of Theoretical Physics, Nuclear Physics
Institute CAS, 25068 \v{R}e\v{z} near Prague, Czech Republic}

\author{Ond{\v r}ej Turek}
\email[]{o.turek@ujf.cas.cz}
\homepage[]{http://researchmap.jp/turek/} \affiliation{Department of Theoretical Physics, Nuclear Physics
Institute CAS, 25068 \v{R}e\v{z} near Prague, Czech Republic,}
\affiliation{Bogoliubov
Laboratory of Theoretical Physics, Joint Institute for Nuclear
Research, 141980 Dubna, Russia,} \affiliation{Laboratory for Unified
Quantum Devices, Kochi University of Technology, Kochi 782-8502,
Japan}

\date{\today}

\begin{abstract}
In contrast to the usual quantum systems which have at most a finite number of open spectral gaps if they are periodic in more than one direction, periodic quantum graphs may have gaps arbitrarily high in the spectrum. This property of graph Hamiltonians, being generic in a sense, inspires the question about the existence of graphs with a finite and nonzero number of spectral gaps. We show that the answer depends on the vertex couplings together with commensurability of the graph edges. A finite and nonzero number of gaps is excluded for graphs with scale invariant couplings; on the other hand, we demonstrate that graphs featuring a finite nonzero number of gaps do exist, illustrating the claim on the example of a rectangular lattice with a suitably tuned $\delta$-coupling at the vertices.
\end{abstract}

 \maketitle

Quantum graphs \cite{BK13} attracted a lot of attention both from
the practical point of view as models of nanostructures
as well as a tool to study properties of quantum systems with a
nontrivial topology of the configuration space. The topological richness of quantum graphs allows them to
exhibit properties different from those of the `usual' quantum
Hamiltonians; examples are well known, for instance, the existence
of compactly supported eigenfunctions on infinite graphs
\cite[Sec.~3.4]{BK13} or the possibility of having flat bands only
as is the case for magnetic chain graphs with a half-of-the-quantum
flux through each chain element \cite{EM15}.

In this letter we are going to consider another situation where
quantum graphs are known to behave unusually. Our problem concerns
the gap structure of the spectrum of periodic quantum graphs. Recall
that the finiteness of the open gap number for periodic quantum
systems in dimension two or more was conjectured in the early days
of quantum theory by Bethe and Sommerfeld \cite{BS33}. The validity of the conjecture was taken for granted even
if its proof turned out to pose a mathematically rather hard problem. It took a long time before it
was rigorously established for the `usual' periodic Schr\"odinger operators
\cite{Sk79,DT82,Sk85,HM98,Pa08}.
Nevertheless, the situation appears to be different for quantum graphs, as we will see below.

The traditional reasoning behind the Bethe-Sommerfeld conjecture relies on
the behavior of the spectral bands identified with the ranges of the
dispersion curves or surfaces which, in contrast to the
one-dimensional situation, typically overlap making opening of gaps
more and more difficult as we proceed to higher energies. The
situation with graphs might be similar \cite[Sec.~4.7]{BK13} but the
spectral behavior need not be the same, one reason being the
possibility of resonant gaps. The existence of gaps coming from a
graph decoration was first observed in the discrete graph
context \cite{SA00} and the effect is present for metric graphs as
well \cite[Sec.~5.1]{BK13}. In addition, the recently discovered
universality property of periodic graphs \cite{BB13} valid in the
generic situation when the graph edges are incommensurate and the
vertex coupling is the simplest possible, usually called Kirchhoff,
shows that the occurrence of infinitely many open gaps is quite
typical.

This prompts one to ask whether there are quantum graphs with the
band spectrum similar to that of the `usual' multidimensional
periodic systems, i.e. a nonzero and finite number of open gaps; for
the sake of brevity we shall speak of the \emph{Bethe--Sommerfeld
property}. With this question in mind, our aim in this letter is twofold. On the one hand we
will show that the answer depends on the vertex coupling  and there
are classes of couplings for which such a behavior is excluded. On the
other hand, using a simple example we are going to demonstrate that
periodic graphs with the Bethe--Sommerfeld property do exist.

Consider an infinite graph $\Gamma$ periodic in $\nu$
directions, with a slight abuse of notation we will speak of a
$\Z^\nu$-periodicity, $\nu\ge 2$. The Hamiltonian is supposed to act
as $-\frac{\D^2}{\D x^2}$ on each edge; to make it a self-adjoint
operator, one has to impose appropriate coupling conditions at each
vertex. The most general form of them \cite{KS99,Ha00} is $(U-I)\Psi
+\mathrm{i}(U+I)\Psi'=0$, where $\Psi,\,\Psi'$ are vectors of the function
and derivative values at the vertex, respectively, and $U$ is a
unitary $n\times n$ matrix for $n$ denoting the degree of the vertex. According
to the eigenvalues of $U$ one can split the coupling into the
Dirichlet, Neumann, and Robin parts \cite[Sec.~1.4]{BK13}.
Equivalently, the coupling can be written in so-called ST-form
\cite{CET10},
\begin{equation}\label{ST}
\left(\begin{array}{cc}
I^{(r)} & T \\
0 & 0
\end{array}\right)
\Psi'= \left(\begin{array}{cc}
S & 0 \\
-T^* & I^{(n-r)}
\end{array}\right)
\Psi
\end{equation}
for certain $r$, $S$, and $T$, where $I^{(r)}$ is the identity
matrix of order $r$ and $S$ is a Hermitian matrix that refers to the Robin part.
A coupling is called \emph{scale-invariant} if the matrix $U$ has no
eigenvalues other than $\pm 1$; it is easy to see that
this happens if and only if $S=0\:$~\cite{CET10b}.  Recall that
the on-shell scattering matrix $\cS(k)$ for the vertex in question
is in the $ST$-formalism given by
$$
\cS(k)=-I^{(n)} +2\begin{pmatrix} I^{(r)} \\ T^*
\end{pmatrix} \left(I^{(r)}+TT^*-\frac{1}{\mathrm{i}k}S\right)^{-1}
\begin{pmatrix} I^{(r)} & T\end{pmatrix}
$$
and it is obvious that $\cS(k)$ is independent of $k$ \emph{iff} $S=0$.

The spectrum is obtained using the Bloch-Floquet theory
\cite[Sec.~4.2]{BK13}. We cut from $\Gamma$ its elementary cell
$\Gamma_\mathrm{per}$ which is assumed to be a finite graph with a
family of pairs of `antipodal' vertices related mutually by the
conditions $\psi(v_+)=\e^{\mathrm{i}\vartheta_l}\psi(v_-)$ and $\psi'(v_+)=\e^{\mathrm{i}\vartheta_l}\psi'(v_-)$ with some $\vartheta_l\in(-\pi,\pi]$, where
$l=1,\ldots,\nu$ with $\nu$ being the dimension of translation group
associated with graph periodicity; the pair of edges with the
endpoints $v_\pm$ can be turned into a single edge by identifying these endpoints,
and the acquired phase $\vartheta_l$ coming from the Bloch
conditions can be also regarded as being induced by a suitable magnetic
potential.

The spectral problem can be solved in the usual way,
cf.~\cite[Sec.~2.1]{BK13} or \cite{BB13}. Assuming that
$\Gamma_\mathrm{per}$ has $E$ edges, we consider three $2E\times2E$
matrices. The diagonal matrix $\mathbf{L}$ is determined by the
lengths of the directed edges (bonds) of $\Gamma_\mathrm{per}$, the
diagonal matrix $\mathbf{A}$ has the entries $\e^{\mathrm{i}\vartheta_l}$ or
$\e^{-\mathrm{i}\vartheta_l}$ at the positions corresponding to the edges
created by the mentioned vertex identification, and all its other
entries are zero, and finally, the matrix $\mathbf{S}$ is the bond
scattering matrix, which contains directed edge-to-edge scattering
coefficients. Using them, we define
\begin{equation} \label{F}
F(k;\vec{\vartheta}):=\det\left(\mathbf{I}
-\e^{\mathrm{i}(\mathbf{A}+k\mathbf{L})}\mathbf{S}(k)\right)\,;
\end{equation}
then $k^2\in\sigma(H)$ holds \emph{iff} there is a
$\vartheta\in(-\pi,\pi]^\nu$ such that the secular equation
$F(k;\vec{\vartheta})=0$ is satisfied.

Suppose now that the couplings at all the vertices of $\Gamma$ are
scale-invariant. This, in particular, means the matrix $\mathbf{S}$
entering formula~(\ref{F}) is independent of $k$, hence the value
$F(k;\vec{\vartheta})$ depends on the vectors $\vec{\vartheta}$ and
$k\ell_0,k\ell_1,\ldots,k\ell_d$, where $\{\ell_0,\ell_1,\ldots,
\ell_d\},\: d\le E-1$, is the set of mutually different edge lengths
of $\Gamma$, and $F(k;\vec{\vartheta})$ is obviously $2\pi$-periodic in the terms
$k\ell_0,k\ell_1,\ldots,k\ell_d$. The secular equation can be then
written as
\begin{equation}\label{secular}
F(\{k\ell_0\}_{(2\pi)},\{k\ell_1\}_{(2\pi)},\ldots,\{k\ell_d\}_{(2\pi)};\vec{\vartheta})=0\,,
\end{equation}
where $\{x\}_{(2\pi)}:= 2\pi\{\frac{x}{2\pi}\}$, for $\,\{\cdot\}$ denoting
the difference between the number and its nearest integer. This allows us to prove that
\begin{enumerate}[(i)]
\setlength{\itemsep}{-3pt}
\item if $\sigma(H)$ has a gap, then it has infinitely many gaps,
\item the gaps can be divided into series with asymptotically constant lengths with respect to
$k$, and
\item in particular, if all the edge lengths are commensurate, the momentum spectrum is periodic.
\end{enumerate}
The easiest part to check is (iii). In that case there is an
elementary length $L>0$ and integers $m_j\in\N$ such that
$\ell_j=m_jL$ holds for $j=0,1,\ldots,d$, and consequently, the
left-hand side of~(\ref{secular}) is $2\pi/L$-periodic with respect
to $k$. Parts (i) and (ii) are more involved and we just sketch the argument
referring to \cite{ET17} for the full proof.

To prove (i) we consider a $k>0$ satisfying $k^2\notin\sigma(H)$ and use it
to prove the existence, for any given $C>0$, a $k'>C$ such that $(k')^2\notin\sigma(H)$.
Due to the continuity of $F$ it is sufficient to find a $k'$ so that
the values $k'\ell_j$ are arbitrarily close to $k\ell_j$ up to an
integer multiple of $2\pi$. To this aim, we denote
$\alpha_j=\frac{\ell_j}{\ell_0}$ and employ the simultaneous version
of the Dirichlet's approximation theorem by which for any $N\in\N$
there are integers $p_1,\ldots,p_d,\:q\in\Z$, $\,1\leq q\leq N$,
such that
\begin{equation}\label{Dirichlet}
\Big|\alpha_j-\frac{p_j}{q}\Big|\leq\frac{1}{qN^{1/d}}\,.
\end{equation}
Choosing integers $m>\frac{\ell_0C}{2\pi}$ and
$N>\left(\frac{2\pi}{\delta}m\right)^d$, and putting
$k'_\delta:=k+2\pi m\frac{q}{\ell_0}$, it is
straightforward to see that $k'_\delta>C$ and, using (\ref{Dirichlet}), to check that
$\big|\left\{k'_\delta\ell_j -k\ell_j\right\}_{(2\pi)}\big|<\delta$ for all $j$.
Moreover, the latter inequality in combination with the argument used to
prove (i) yields the claim (ii).

In fact, one can exclude the Bethe--Sommerfeld property for a wider
class of graphs. Given a vertex coupling described by
condition~(\ref{ST}), we consider the \emph{associated
scale-invariant} one obtained by replacing the Robin part $S$ by
zero. The vertex scattering matrix can be then written as
$\cS(k)=\cS_0+\frac{1}{k}\cS_1(k)$, where $\cS_0$ is the scattering matrix of the associated scale-invariant vertex coupling. Then the function
$F(k;\vec{\vartheta})$ in the secular equation is
$$
F_0(\{k\ell_0\}_{(2\pi)},\{k\ell_1\}_{(2\pi)},\ldots,\{k\ell_d\}_{(2\pi)};\vec{\vartheta})
+\frac{1}{k}F_1(k;\vec{\vartheta})\,,
$$
where the subscript zero at $F_0$ refers to the Hamiltonian $H_0$ of the graph in which
all the couplings have been replaced by the associated
scale-invariant ones. Using the fact that the leading behavior of
$F(\cdot;\vec{\vartheta})$ at high energies comes from the
scale-invariant term, it is not difficult to see that \cite{ET17}
\begin{enumerate}[(i)]
\setlength{\itemsep}{-3pt}
\item if $\sigma(H_0)$ has an open gap, then $\sigma(H)$ has infinitely many
gaps,
\item if all the edge lengths of $\Gamma$ are commensurate, then the gaps of $\sigma(H)$
asymptotically coincide with those of $\sigma(H_0)$.
\end{enumerate}

\begin{figure}[tbp]
	\centering
	\hspace{0em}\includegraphics[width=.5\columnwidth]{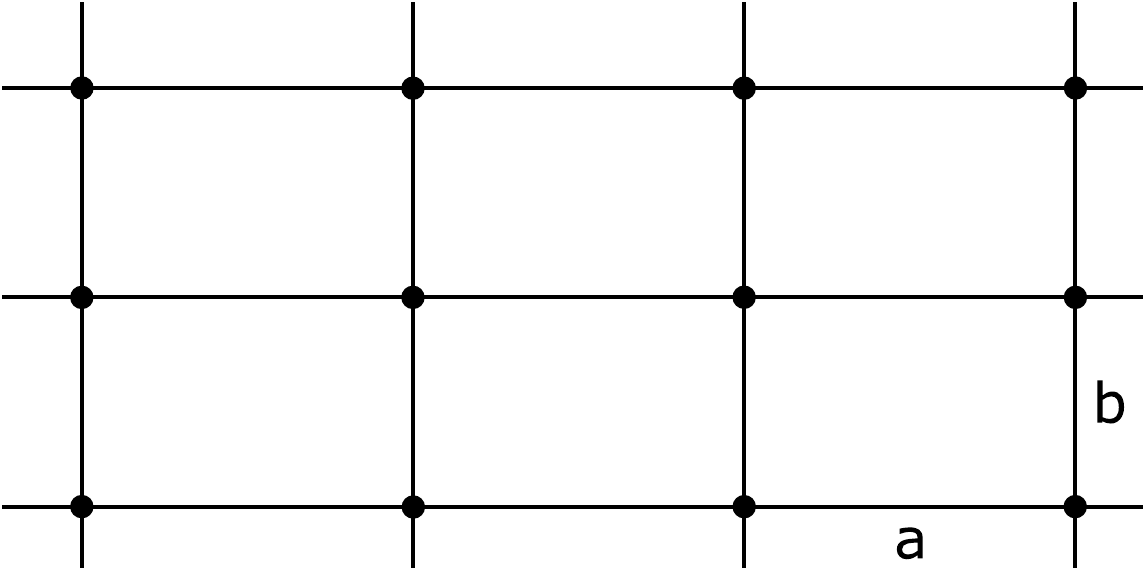}
	\caption{The rectangular-lattice graph}
	\label{fig:lattice}
\end{figure}

Let us turn to our second topic, the existence of graphs with the
Bethe-Sommerfeld property. To this goal we revisit the \emph{lattice
Kronig-Penney model} introduced in \cite{Ex95} and further discussed
in \cite{Ex96, EG96}. In this case $\Gamma$ is a rectangular lattice
graph in the plane with edges of lengths $a$ and $b$, cf.~Fig.~\ref{fig:lattice}.
The Hamiltonian
$H=H_{\alpha,a/b}$ is the negative Laplacian with the $\delta$ coupling
condition in each vertex, i.e. the functions are continuous there
and satisfy $\sum_{j=1}^4 \psi'(v) = \alpha\psi(v)$ with a fixed
parameter $\alpha\in\R$.

According to \cite{Ex96}, a number $k^2>0$ belongs to a gap if and
only if $k>0$ satisfies the gap condition, which reads
\begin{equation}\label{alpha>0}
\tan\left(\frac{ka}{2}-\frac{\pi}{2}\left\lfloor\frac{ka}{\pi}\right\rfloor\right)
+\tan\left(\frac{kb}{2}-\frac{\pi}{2}\left\lfloor\frac{kb}{\pi}\right\rfloor\right)<\frac{|\alpha|}{2k}
\end{equation}
for $\alpha>0$ and the analogous one with $\tan$ replaced by $\cot$
if $\alpha<0$; we neglect the case $\alpha=0$ where the spectrum is
trivial, $\sigma(H)=[0,\infty)$. For $\alpha<0$ the spectrum extends
to the negative part of the real axis and may have a gap there, but
since such a gap always has a positive part \cite{EG96}, we may restrict our attention to examining gaps in the positive spectrum.

The crucial quantity is the ratio $\theta=\frac{a}{b}$. It is
obvious that $\sigma(H)$ has infinitely many gaps once $\alpha\ne 0$
and $\theta$ is rational, and the same is true for the `well approximable'
irrationals \cite{Ex96}. We thus focus on the other irrationals, called
\emph{badly approximable}, i.e those to which there is a $c>0$ such
that
$$
\Big|\theta-\frac{p}{q}\Big|>\frac{c}{q^2}
$$
for all $p,q\in\Z$ with $q\neq0$. These numbers form a set of zero
Lebesgue measure. Alternatively they can be characterized as irrationals
for which the sequence in the continued-fraction representation,
$\theta=[c_0;c_1,c_2,\ldots]$, is bounded \cite{Kh64}, or as numbers whose
\emph{Markov constant} $\mu(\theta)$, defined~\cite{Ca57} for
$\theta\in\mathbb{R}$ as
\begin{equation}\label{mu(theta)}
\mu(\theta)=\inf\Big\{c>0\;\Big|\;\Big(\exists_\infty(p,q)\in\mathbb{N}^2\Big)
\Big(\Big|\theta-\frac{p}{q}\Big|<\frac{c}{q^2}\Big)\Big\},
\end{equation}
is strictly positive.
It is convenient to introduce a one-sided analogue $\upsilon(\theta)$ of the Markov constant, with the last inequality in (\ref{mu(theta)}) replaced by $0<\theta-\frac{p}{q}<\frac{c}{q^2}$. We have $\mu(\theta)=\min\{\upsilon(\theta),\upsilon(\theta^{-1})\}$; the number $\upsilon(\theta)$ may or may not coincide with the Markov constant \cite{PSZ16}.

To get the existence claim we focus on the situation where $\theta$ is the `worst approximable' irrational, the \emph{golden mean}, $\phi=\frac{\sqrt{5}+1}{2}$ .
Our result about golden-mean lattice is the following:
\begin{enumerate}[(i)]
\setlength{\itemsep}{-3pt}
\item If $\alpha>\frac{\pi^2}{\sqrt{5}a}$ or $\alpha\leq-\frac{\pi^2}{\sqrt{5}a}$,
the spectrum has infinitely many gaps.
\item If
$-\frac{2\pi}{a}\tan\left(\frac{3-\sqrt{5}}{4}\pi\right)\leq\alpha\leq\frac{\pi^2}{\sqrt{5}a}$
there are no gaps in the spectrum.
\item If
$-\frac{\pi^2}{\sqrt{5}a}<\alpha<-\frac{2\pi}{a}\tan\left(\frac{3-\sqrt{5}}{4}\pi\right)$,
there is a nonzero and finite number of gaps in the
spectrum.
\item Moreover, put $A_{j}:=\frac{2\pi\left(\phi^{2j}-\phi^{-2j}\right)}
{\sqrt{5}}\tan\left(\frac{\pi}{2}\phi^{-2j}\right)$, then there are
exactly $N$ gaps in the spectrum if
$-A_{N+1}\leq\alpha<-A_N$.
\end{enumerate}
Note
that the window in which Bethe--Sommerfeld property occurs in
this example (statement (iii)), is rather narrow, roughly $4.298 \lesssim -\alpha a
\lesssim 4.414$.

The proof of these claims is rather involved and we limit ourselves with mentioning its key elements referring to \cite{ET17} for the full exposition.
The central notion is that of the \emph{Diophantine approximation of third type from below} (\emph{from above}, respectively). The former is a number $\frac{p}{q}$ with $p,q\in\mathbb{Z}$ such that
\begin{equation}\label{third below}
0<q(q\theta-p)<q'(q'\theta-p')
\end{equation}
holds for all $\frac{p'}{q'}\geq\theta$ with $\frac{p'}{q'}\neq\frac{p}{q}$, $p',q'\in\mathbb{Z}$ and $0<q'\leq q$; the approximation from above has (\ref{third below}) replaced by $0<q(p-q\theta)<q'(p'-q'\theta)$. We note that $\upsilon(\theta)$ is the infimum of those $q(q\theta-p)$ for which $\frac{p}{q}$ is a best approximation from below to $\theta$. These approximations are also closely related to \emph{convergents} obtained from truncated continued-fraction representation of $\theta$. Specifically, every best approximation of the third kind from below to a $\theta\in\mathbb{R}$ is a convergent of $\theta$, and on the other hand, every best approximation from above is either $\lceil\theta\rceil$ or a convergent of $\theta$, where $\lceil\cdot\rceil$ is the ceiling function.

The described Diophantine approximation in combination with the gap condition allows us to estimate the number of gaps for a given ratio $\theta=a/b$ and coupling parameter $\alpha$. This has to be done for each sign of $\alpha$ separately. If $\alpha>0$, condition (\ref{alpha>0}) yields that
\begin{itemize}
\item if $\alpha<\pi^2\cdot\min\left\{\frac{\upsilon(\theta)}{b},\frac{\upsilon(\theta^{-1})}{a}\right\}$, the spectrum has at most finitely many gaps. If the opposite (sharp) inequality holds true, the number of gaps is infinite;
\item if $\alpha\leq\gamma_+$ for $\gamma_+$ given by
\begin{equation}\label{gamma}
\gamma_+:=\min_{\eta=\theta,\theta^{-1}} \inf_{m\in\mathbb{N}} \left\{2\pi m\sqrt{\frac{\eta}{ab}}\tan\left(\frac{\pi}{2}(m\eta-\lfloor m\eta\rfloor)\right)\right\},
\end{equation}
the spectrum has no gaps. If $\alpha>\gamma_+$, there are gaps in the spectrum;
\item in particular, if $\gamma_+<\alpha<\pi^2\cdot\min\left\{\frac{\upsilon(\theta)}{b},\frac{\upsilon(\theta^{-1})}{a}\right\}$,
there is a nonzero and finite number of gaps in the spectrum.
\end{itemize}
The golden mean $\phi$ in our example has continued-fraction representation $\phi=[1;1,1,\dots]$, therefore, its convergents are ratios $\frac{F_{n+1}}{F_n}$ of the Fibonnaci numbers $F_n=\frac{\phi^n-(-\phi)^{-n}}{\sqrt{5}}$, and we have $\mu(\phi)=\upsilon(\phi)=\frac{1}{\sqrt{5}}$ in view of the Hurwitz theorem \cite{Hu1891}.
Hence we get
$$
\gamma_+=\frac{\pi^2}{\sqrt{5}a} \quad\text{and}\quad \pi^2\cdot\min\left\{\frac{\upsilon(\phi)}{b},\frac{\upsilon(\phi^{-1})}{a}\right\}=\frac{\pi^2}{\sqrt{5}a}\,,
$$
which implies for all positive $\alpha$
either infinite number of spectral gaps or none at all. On the other hand, for $\alpha<0$ the gap condition implies that
\begin{itemize}
\item if $|\alpha|<\pi^2\cdot\min\left\{\frac{\upsilon(\theta)}{a},\frac{\upsilon(\theta^{-1})}{b}\right\}$, the number of gaps in the positive spectrum is at most finite. By contrast, for $|\alpha|$ greater than the right-hand side of the above inequality, there are infinitely many spectral gaps, while
\item if $|\alpha|\leq\gamma_-$ for $\gamma_-$ given by the relation analogous to (\ref{gamma}) in which
$m\eta-\lfloor m\eta\rfloor$ is replaced by $\lceil m\eta\rceil-m\eta$,
the spectrum has no gaps. If $|\alpha|>\gamma_-$, there are gaps in the spectrum;
\item in particular, if $\gamma_-<|\alpha|<\pi^2\cdot\min\left\{\frac{\upsilon(\theta)}{a},\frac{\upsilon(\theta^{-1})}{b}\right\}$,
there is a nonzero and finite number of gaps in the spectrum.
\end{itemize}
The last one of these results together with the easily verifiable formula
$$
\gamma_-=\frac{2\pi}{a}\tan\frac{(3-\sqrt{5})\pi}{4}
$$
proves the claim (iii) above, demonstrating thus the existence of graphs with the Bethe--Sommerfeld property. To prove (iv), the first step consists in showing that the number of gaps in the golden-mean lattice graph is equal to the number of solutions $m\in\mathbb{N}$ of
$$
\frac{2\pi m}{a}\tan\left(\frac{\pi}{2}\left(\lceil m\phi\rceil-m\phi\right)\right)<|\alpha|\,.
$$
If $-A_{N+1}\leq\alpha<-A_N$, one can check that the inequality is satisfied only for $m=F_n$ with $n=2,4,6,\ldots,2N$; this implies the existence of exactly $N$ gaps.

Since a finite nonzero number of gaps occurred in the above example only for attractive vertex couplings, it is natural to ask whether the attractivity of the coupling is always a necessary condition for the Bethe--Sommerfeld property. It appears that it is not the case, a more thorough analysis of the gap condition \cite{ET17} shows that, for instance, the edge ratio
$$
\theta=\frac{2t^3-2t^2-1+\sqrt{5}}{2(t^4-t^3+t^2-t+1)} \quad\mathrm{with}\;\, t\in\mathbb{N},\; t\geq3\,,
$$
which has the continued-fraction representation $[0;t,t,1,1,1,1,\ldots]$, yields the lattice graph spectrum with the said property for a certain $\alpha>0$ and for a certain $\alpha<0$ as well. This observation can be stated more generally~\cite{ET17} and allows to explicitly construct ratios to achieve the Bethe--Sommerfeld property.

In conclusion, we have proved and demonstrated on concrete examples that there are periodic quantum graphs the spectrum of which contains a nonzero and finite number of open gaps. We have also described how this property depends on the type of the vertex coupling; in particular, we showed that a quantum graph cannot be of the Bethe--Sommerfeld type if its couplings are scale invariant or associated to scale-invariant ones.

\smallskip

The research was supported by the Czech Science Foundation
(GA\v{C}R) within the project 17-01706S.

\end{document}